\documentclass[copyright,creativecommons]{eptcs}

\pdfoutput=1

\providecommand{\event}{FROM 2022}

\usepackage[T1]{fontenc}

\usepackage{amsmath}
\usepackage{amsthm}

\usepackage{newtxtext}
\usepackage{newtxmath}

\usepackage{tikz}
\usetikzlibrary{petri}
\usetikzlibrary{arrows.meta}

\newcommand{\deadlock}{\mathbf{0}} 
\newcommand{\send}{\lhd}
\newcommand{\receive}[1]{\rhd#1.\>}
\newcommand{\newchannel}[1]{\nu#1.\>}
\newcommand{\bridge}{\mathbin\rightarrow}
\newcommand{\distributor}{\mathbin\Rightarrow}
\newcommand{\unreliability}[1]{\text{\textup{\textcurrency}}^{#1}}
\newcommand{\loser}{\unreliability{?}}
\newcommand{\duplicator}{\unreliability{+}}
\newcommand{\duploser}{\unreliability{*}}
\newcommand{\Parallel}[2]{{\textstyle\prod{}}#1\mathbin\leftarrow#2.\>}

\tikzset{
    communication net/.style={
        scale=1.3,
        every place/.style={minimum size=5.5mm,semithick},
        every transition/.style={minimum size=3.5mm,semithick},
        pre/.style={
            <-,
            shorten <=1pt,
            >={Straight Barb[angle=90:2pt]},
            semithick
        },
        post/.style={
            ->,
            shorten >=1pt,
            >={Straight Barb[angle=90:2pt]},
            semithick
        },
        local/.style={double},
        symbol/.style={anchor=center},
        captions/.style={
            yshift=-0.9cm,
            every node/.style={anchor=base}
        }
    }
}

\newcommand{\losing}{$?$}
\newcommand{\duplicating}{$+$}
\newcommand{\duplosing}{$*$}

\newcommand{\verticaldots}{%
    \makebox[0cm][l]{\raisebox{0.4cm}{$\cdot$}}%
    \makebox[0cm][l]{\raisebox{-0.4cm}{$\cdot$}}%
    $\cdot$%
}

\theoremstyle{definition}
\newtheorem{definition}{Definition}

\title{
    Correctness of Broadcast via Multicast:\\
    Graphically and Formally
}
\newcommand{\titlerunning}{%
    Correctness of Broadcast via Multicast:
    Graphically and Formally%
}
\author{
    Wolfgang Jeltsch
    \institute{Well-Typed\\London, England}
    \email{wolfgang@well-typed.com}
\and
    Javier D\'iaz
    \institute{Atix Labs (a Globant Division)\\Buenos Aires, Argentina}
    \email{javier.diaz@globant.com}
}
\newcommand{\authorrunning}{Wolfgang Jeltsch \& Javier D\'iaz}
\newcommand{\copyrightholders}{Input Output}

\begin{document}

\maketitle

\begin{abstract}

Maintaining data consistency among multiple parties requires nodes to
repeatedly send data to all other nodes. For example, the nodes of a
blockchain network have to disseminate the blocks they create across the
whole network. The scientific literature typically takes the ideal
perspective that such data distribution is performed by broadcasting to
all nodes directly, while in practice data is distributed by repeated
multicast. Since correctness and security of consistency maintenance
protocols usually have been established for the ideal setting only, it
is vital to show that these properties carry over to real-world
implementations. Therefore, it is desirable to prove that the ideal and
the real behavior are equivalent.

In the work described in this paper, we take an important step towards
such a proof by proving a simpler variant of this equivalence statement.
The simplification is that we consider only a concrete pair of network
topologies, which nevertheless illustrates important phenomena
encountered with arbitrary topologies. For describing systems that
distribute data, we use a domain-specific language of processes that
corresponds to a class of Petri nets and is embedded in a
general-purpose process calculus. This way, we can outline our proof
using an intuitive graphical notation and leverage the rich theory of
process calculi in the actual proof, which is machine-checked using the
Isabelle proof assistant.

\end{abstract}

\section{Introduction}

Systems that maintain data consistency among multiple parties are
becoming increasingly relevant. For example, blockchains have seen
growing utilization in areas such as finance, identification, logistics,
and real estate. Incorrect behavior of such systems may often result in
serious damage. It is therefore valuable to formally prove their
correctness and security.

Keeping data consistent requires nodes to regularly disseminate data to
other nodes. A perspective often taken in the scientific literature is
that such data is broadcast to all nodes directly. For example, the
descriptions of the blockchain consensus protocols of the Ouroboros
family~\cite{badertscher:2018,david:2018,kiayias:2017} assume such
direct communication. In practice, however, data is distributed via
repeated multicast. Since correctness and security typically have been
established for the ideal setting only, it is vital to show that they
carry over to real-world implementations.

In this paper, we take an important step in this direction by showing
that the ideal behavior of direct broadcast and the real behavior of
broadcast via multicast are in a certain sense equivalent. Concretely,
we make the following contributions:
\begin{itemize}

\item

In Sect.~\ref{section:communication-language}, we define a restricted
language of processes that are able to describe network communication.
Processes in our language correspond to hierarchical Petri nets with
exactly one input place per transition. We define our language via an
embedding in a general-purpose process calculus. This approach enables
us to leverage the rich theory of process calculi while allowing us to
use an intuitive graphical notation similar to the one of Petri nets.

\item

In Sect.~\ref{section:behavioral-equivalence}, we devise a notion of
behavioral equivalence of networks that does not distinguish between
different patterns of packet arrival. Our approach is to start with
bisimilarity and weaken it by amending the involved processes to allow
for additional behavior. Building on bisimilarity permits us to reason
in a modular fashion.

\item

In Sect.~\ref{section:correctness-proof}, we present a proof of
behavioral equivalence of broadcast via multicast and direct broadcast
under the assumption that network communication may involve packet loss
and duplication. Our proof is about a concrete pair of networks that
nevertheless captures important general phenomena. The proof works by
rewriting a process describing the former form of broadcast into a
process describing the latter. For the individual rewriting steps, we
rely on certain fundamental lemmas, not all of which have been proved so
far. We outline our proof using the graphical notation for processes and
show the proof's first part in detail. The whole proof is formalized in
Isabelle/HOL.

\end{itemize}
Afterwards, we discuss related work in Sect.~\ref{section:related-work}
and give a conclusion and an outlook on ongoing and future work in
Sects.~\ref{section:conclusion} and
\ref{section:ongoing-and-future-work}.

\section{A Language for Communication Networks}

\label{section:communication-language}

For describing communication networks, we use a custom language of
processes that communicate via asynchronous channels. Let uppercase
letters denote processes and lowercase letters denote channels. The
syntax of our communication language is given by the following BNF rule:
\begin{equation*}
\mathit{Process} \mathrel{{\mathop:}{\mathop:}{=}}
    \deadlock                             \mid
    P \parallel Q                         \mid
    \newchannel{a} P                      \mid
    a \bridge b                           \mid
    a \distributor [b_{1}, \ldots, b_{n}] \mid
    \loser a                              \mid
    \duplicator a                         \mid
    \duploser a
\end{equation*}
A process is one of the following:
\begin{itemize}

\item

The \emph{stop process}~$\deadlock$, which does nothing

\item

A \emph{parallel composition} $P \parallel Q$, which performs
$P$~and~$Q$ in parallel

\item

A \emph{restricted process} $\newchannel{a} P$, which behaves like~$P$
except that the channel~$a$ is local

\item

A \emph{bridge} $a \bridge b$, which continuously forwards packets from
channel~$a$ to channel~$b$

\item

A \emph{distributor} $a \distributor [b_{1}, \ldots, b_{n}]$, which
continuously forwards packets from channel~$a$ to all channels~$b_{i}$

\item

A \emph{loser} $\loser a$, which continuously drops packets from
channel~$a$

\item

A \emph{duplicator} $\duplicator a$, which continuously duplicates
packets in channel~$a$

\item

A \emph{duploser} $\duploser a$, which continuously drops packets from
and continuously duplicates packets in channel~$a$

\end{itemize}
Parallel composition has lowest precedence, restriction has intermediate
precedence, and all other constructs have highest precedence.

Our communication language is embedded in the \TH-calculus\footnote{See
\url{https://github.com/input-output-hk/thorn-calculus}.} (pronounced
``thorn calculus''), a general-purpose process calculus that is itself
embedded in Isabelle/HOL. Parallel composition, restriction, and the
stop process are in fact constructs of the \TH-calculus, which we
directly use in the communication language. Besides these three
constructs, the \TH-calculus provides constructs for sending and
receiving: $a \send x$ is a process that sends value~$x$ to channel~$a$,
and $a \receive{x} P$ is a process that receives a value~$x$ from
channel~$a$ and continues like~$P$, where $P$ may mention~$x$. The
operational semantics of the \TH-calculus is essentially the one of the
asynchronous $\pi$-calculus~\cite{honda:1991}.

The constructs of our communication language other than those that stem
directly from the \TH-calculus are defined in terms of \TH-calculus
constructs as follows:
\begin{align}
a \distributor [b_{1}, \ldots, b_{n}]
& =
a \receive{x} (b_{1} \send x \parallel \ldots \parallel b_{n} \send x
\parallel a \distributor [b_{1}, \ldots, b_{n}])
\\
\label{equation:bridge}
a \bridge b
& =
a \distributor [b]
\\
\loser a
& =
a \distributor []
\\
\duplicator a
& =
a \distributor [a, a]
\\
\duploser a
& =
\loser a \parallel \duplicator a
\end{align}

Note that distributors with a common source channel compete for the
packets in that channel. As a consequence, the presence of a loser or
duplicator for some channel does not mean that all packets sent to this
channel are lost or are duplicated forever, because other distributors
might get hold of packets in this channel and thus prevent the loser or
duplicator from fetching them. Furthermore, observe that, while the
\TH-calculus has the mobility feature of the $\pi$-calculus, this
feature cannot be exploited by processes of the communication language,
since such processes cannot specifically send and receive channels, but
only forward data, treating it as black boxes.

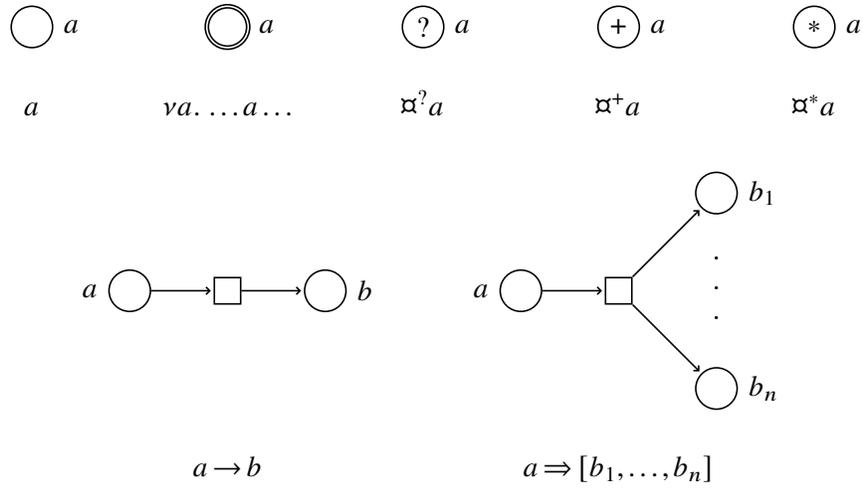
\begin{figure}

\centering
\begin{tikzpicture}[communication net]

\begin{scope}[example channel/.style={place,label=right:$a$}]

\node [example channel] at (-4,0)         {};
\node [example channel] at (-2,0) [local] {};
\node [example channel] at ( 0,0)         {\losing};
\node [example channel] at ( 2,0)         {\duplicating};
\node [example channel] at ( 4,0)         {\duplosing};

\end{scope}

\begin{scope}[captions]

\node at (-4,0) {$a$};
\node at (-2,0) {$\newchannel{a} \ldots a \ldots$};
\node at ( 0,0) {$\loser a$};
\node at ( 2,0) {$\duplicator a$};
\node at ( 4,0) {$\duploser a$};

\end{scope}

\begin{scope}[yshift=-3.7cm]

\begin{scope}

\node [place] (a) at (-3,1) [label=left:$a$]  {};
\node [place] (b) at (-1,1) [label=right:$b$] {};

\node [transition] at (-2,1) {} edge [pre]  (a)
                                edge [post] (b);

\end{scope}

\begin{scope}

\node [place] (a)  at (1,1) [label=left:$a$]      {};
\node [place] (b1) at (3,2) [label=right:$b_{1}$] {};
\node [place] (bn) at (3,0) [label=right:$b_{n}$] {};

\node [symbol] at (3,1) {\verticaldots};

\node [transition] at (2,1) {} edge [pre]  (a)
                               edge [post] (b1)
                               edge [post] (bn);

\end{scope}

\begin{scope}[captions]

\node at (-2,0) {$a \bridge b$};
\node at ( 2,0) {$a \distributor [b_{1}, \ldots, b_{n}]$};

\end{scope}

\end{scope}

\end{tikzpicture}

\caption{Elements of communication nets}

\label{figure:communication-nets}

\end{figure}

The sublanguage formed by $\deadlock$, $\parallel$, $\nu$, and
$\distributor$ corresponds to hierarchical Petri nets with
exactly one input place per transition, with the constructs of the two
formalisms corresponding to each other as follows:
\begingroup
\newcommand{\correspondence}[2]{\text{#1}&\leftrightarrow\text{#2}}
\begin{align*}
\correspondence
    {channel}
    {place}
\\
\correspondence
    {distributor}
    {transition}
\\
\correspondence
    {stop process}
    {Petri net without transitions}
\\
\correspondence
    {parallel composition}
    {Petri net join that identifies equal places}
\\
\correspondence
    {restricted process}
    {subnet}
\end{align*}
\endgroup
This correspondence allows us to use the graphical notation for
hierarchical Petri nets to depict processes of that fragment of the
communication language. We modify the representation of restricted
processes and add custom notation for the remaining constructs, leading
to the graphical notation shown in Fig.~\ref{figure:communication-nets}.
We call diagrams that use this notation \emph{communication nets}.

\begin{figure}

\centering
\begin{tikzpicture}[communication net]

\begin{scope}

\node [place] (a) at (-4,0) [label=left:$a$] {};

\node [transition] at (-3,0) {} edge [pre] (a);

\end{scope}

\begin{scope}

\node [place] (a) at (-1,0) [label=left:$a$] {};

\node [transition] at (0,0) {} edge [pre]             (a)
                               edge [post,bend left]  (a)
                               edge [post,bend right] (a);

\end{scope}

\begin{scope}

\node [place] (a) at (3,0) [label=above:$a$] {};

\node [transition] at (2,0) {} edge [pre]             (a);
\node [transition] at (4,0) {} edge [pre]             (a)
                               edge [post,bend left]  (a)
                               edge [post,bend right] (a);

\end{scope}

\begin{scope}[captions]

\node at (-3.5,0) {$\loser a$};
\node at (-0.5,0) {$\duplicator a$};
\node at (   3,0) {$\duploser a$};

\end{scope}

\end{tikzpicture}

\caption{%
    Explicit communication net representations
    of unreliability constructs%
}

\label{figure:unreliability-explicitly}

\end{figure}
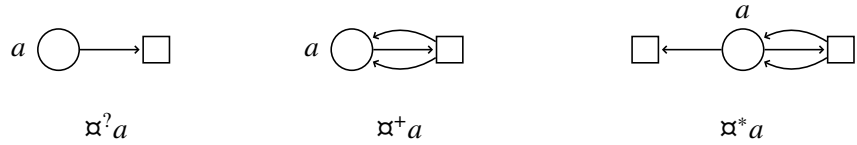

Observe that the notation for bridges just reflects their definition in
terms of distributors, as shown in (\ref{equation:bridge}), and
therefore actually does not constitute an extension to Petri net
notation. The notation for losers, duplicators, and duplosers is really
an extension, but since these constructs are defined in terms of
distributors as well, their notation is only syntactic sugar, which we
could drop in favor of the explicit representations shown in
Fig.~\ref{figure:unreliability-explicitly}.

Communication nets identify bisimilar (and thus weakly bisimilar)
processes to some degree. This can ease reasoning, as it means that
certain transformations of processes into bisimilar ones become no-ops
when working on the more abstract level of communication nets.
Concretely, the following bisimilarities are implicit in the
communication net notation:
\begin{align}
\deadlock \parallel P
& \sim
P
\\
P \parallel \deadlock
& \sim
P
\\
(P \parallel Q) \parallel R
& \sim
P \parallel (Q \parallel R)
\\
P \parallel Q
& \sim
Q \parallel P
\\
\label{equation:scope-extension}
P \parallel \newchannel{a} Q
& \sim
\newchannel{a} (P \parallel Q)
&& \hspace{-4em}
\text{if $a$ does not occur freely in~$P$}
\\
\newchannel{a} \newchannel{b} P
& \sim
\newchannel{b} \newchannel{a} P
\\
\label{equation:scope-redundancy}
\newchannel{a} P
& \sim
P
&& \hspace{-4em}
\text{if $a$ does not occur freely in~$P$}
\end{align}
The reason for (\ref{equation:scope-extension}) to
(\ref{equation:scope-redundancy}) being implicit is that the scopes of
local channels are not reflected by communication nets, which just show
locality as a property of channels. Actually,
(\ref{equation:scope-redundancy}) is not implicit in general, only when
the unused channel~$a$ is not depicted, which should be the norm, but is
not required.

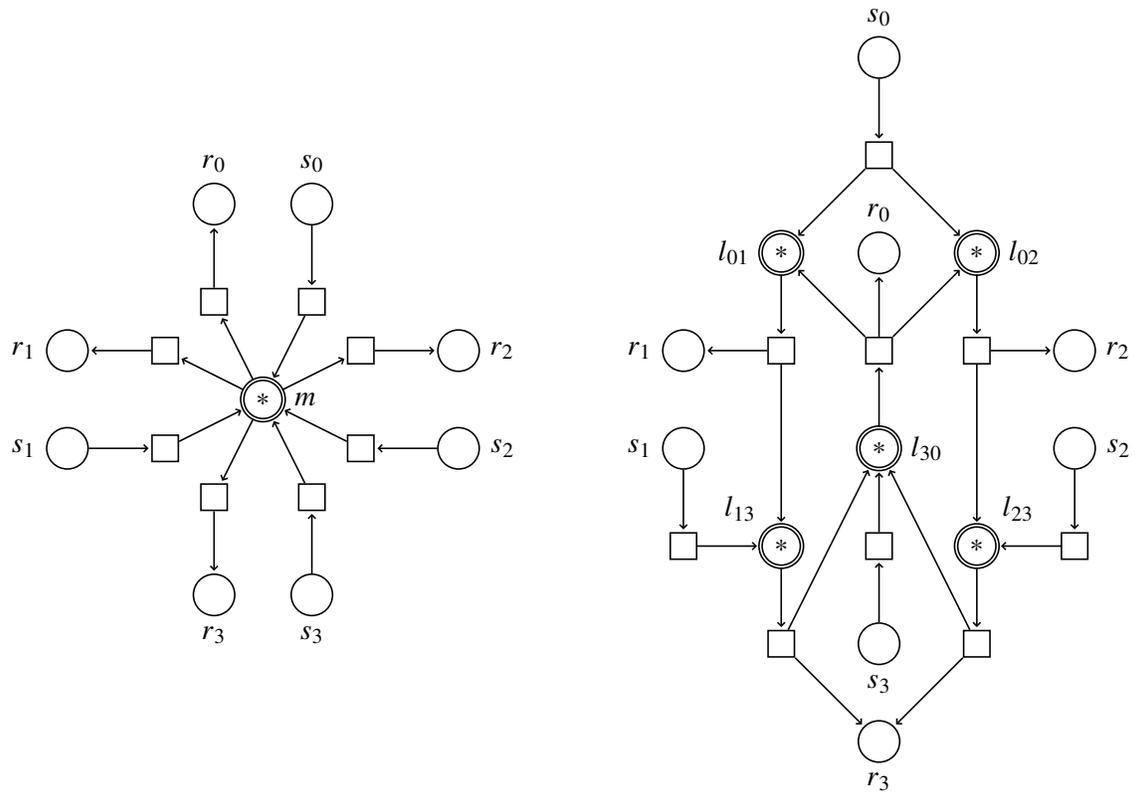
\begin{figure}

\centering
\begin{tikzpicture}[communication net]

\begin{scope}

\node [place] (s0) at ( 0.5,   2) [label=above:$s_{0}$]   {};
\node [place] (s1) at (  -2,-0.5) [label=left:$s_{1}$]    {};
\node [place] (s2) at (   2,-0.5) [label=right:$s_{2}$]   {};
\node [place] (s3) at ( 0.5,  -2) [label=below:$s_{3}$]   {};
\node [place] (r0) at (-0.5,   2) [label=above:$r_{0}$]   {};
\node [place] (r1) at (  -2, 0.5) [label=left:$r_{1}$]    {};
\node [place] (r2) at (   2, 0.5) [label=right:$r_{2}$]   {};
\node [place] (r3) at (-0.5,  -2) [label=below:$r_{3}$]   {};
\node [place] (m)  at (   0,   0) [local,label=right:$m$] {\duplosing};

\node [transition] at ( 0.5,   1) {} edge [pre]  (s0)
                                     edge [post] (m);
\node [transition] at (  -1,-0.5) {} edge [pre]  (s1)
                                     edge [post] (m);
\node [transition] at (   1,-0.5) {} edge [pre]  (s2)
                                     edge [post] (m);
\node [transition] at ( 0.5,  -1) {} edge [pre]  (s3)
                                     edge [post] (m);
\node [transition] at (-0.5,   1) {} edge [pre]  (m)
                                     edge [post] (r0);
\node [transition] at (  -1, 0.5) {} edge [pre]  (m)
                                     edge [post] (r1);
\node [transition] at (   1, 0.5) {} edge [pre]  (m)
                                     edge [post] (r2);
\node [transition] at (-0.5,  -1) {} edge [pre]  (m)
                                     edge [post] (r3);

\end{scope}

\begin{scope}[xshift=6.3cm,yshift=-0.5cm]

\newcommand{\dl}{\duplosing}

\node [place] (s0)  at ( 0, 4) [label=above:$s_{0}$]              {};
\node [place] (s1)  at (-2, 0) [label=left:$s_{1}$]               {};
\node [place] (s2)  at ( 2, 0) [label=right:$s_{2}$]              {};
\node [place] (s3)  at ( 0,-2) [label=below:$s_{3}$]              {};
\node [place] (r0)  at ( 0, 2) [label=above:$r_{0}$]              {};
\node [place] (r1)  at (-2, 1) [label=left:$r_{1}$]               {};
\node [place] (r2)  at ( 2, 1) [label=right:$r_{2}$]              {};
\node [place] (r3)  at ( 0,-3) [label=below:$r_{3}$]              {};
\node [place] (l01) at (-1, 2) [local,label=left:$l_{01}$]        {\dl};
\node [place] (l02) at ( 1, 2) [local,label=right:$l_{02}$]       {\dl};
\node [place] (l13) at (-1,-1) [local,label=above left:$l_{13}$]  {\dl};
\node [place] (l23) at ( 1,-1) [local,label=above right:$l_{23}$] {\dl};
\node [place] (l30) at ( 0, 0) [local,label=right:$l_{30}$]       {\dl};

\node [transition] at ( 0, 3) {} edge [pre]  (s0)
                                 edge [post] (l01)
                                 edge [post] (l02);
\node [transition] at (-2,-1) {} edge [pre]  (s1)
                                 edge [post] (l13);
\node [transition] at ( 2,-1) {} edge [pre]  (s2)
                                 edge [post] (l23);
\node [transition] at ( 0,-1) {} edge [pre]  (s3)
                                 edge [post] (l30);
\node [transition] at (-1, 1) {} edge [pre]  (l01)
                                 edge [post] (r1)
                                 edge [post] (l13);
\node [transition] at ( 1, 1) {} edge [pre]  (l02)
                                 edge [post] (r2)
                                 edge [post] (l23);
\node [transition] at (-1,-2) {} edge [pre]  (l13)
                                 edge [post] (r3)
                                 edge [post] (l30);
\node [transition] at ( 1,-2) {} edge [pre]  (l23)
                                 edge [post] (r3)
                                 edge [post] (l30);
\node [transition] at ( 0, 1) {} edge [pre]  (l30)
                                 edge [post] (r0)
                                 edge [post] (l01)
                                 edge [post] (l02);

\end{scope}

\end{tikzpicture}

\caption{%
    Example of direct broadcast (left)
    and broadcast via multicast (right)%
}

\label{figure:broadcast}

\end{figure}

Two complete communication nets are shown in
Fig.~\ref{figure:broadcast}. They characterize the two networks whose
behavioral equivalence we show in Sect.~\ref{section:correctness-proof}.
In both communication nets, channels $s_{i}$ and $r_{i}$ form the
interface of a network node~$i$, with $s_{i}$ accepting packets for
sending and $r_{i}$ providing received packets. The local channel~$m$ in
the left communication net represents the broadcast medium, and each
local channel $l_{ij}$ in the right communication net represents a
multicast link from node~$i$ to node~$j$. Note that we assume any
communication to be unreliable, which is reflected by the channel~$m$
and all channels~$l_{ij}$ having duplosers attached. The duplicator part
of the duploser attached to~$m$ has the additional purpose of allowing
packets to reach all nodes instead of only one node each.

The two communication nets in Fig.~\ref{figure:broadcast} correspond to
processes~$D$ (direct broadcast) and~$M$ (broadcast via multicast)
defined as follows:
\begin{align}
\label{equation:direct-broadcast}
D & =
\newchannel{m}
(
    \duploser m
    \parallel s_{0} \bridge m
    \parallel s_{1} \bridge m
    \parallel s_{2} \bridge m
    \parallel s_{3} \bridge m
    \parallel m \bridge r_{0}
    \parallel m \bridge r_{1}
    \parallel m \bridge r_{2}
    \parallel m \bridge r_{3}
)
\\
\label{equation:broadcast-via-multicast}
M & =
\newchannel{l_{01}}
\newchannel{l_{02}}
\newchannel{l_{13}}
\newchannel{l_{23}}
\newchannel{l_{30}}
(M_{*} \parallel M_{\mathrm{i}} \parallel M_{\mathrm{o}})
\\
M_{*} & =
\duploser l_{01} \parallel
\duploser l_{02} \parallel
\duploser l_{13} \parallel
\duploser l_{23} \parallel
\duploser l_{30}
\\
M_{\mathrm{i}} & =
s_{0} \distributor [l_{01}, l_{02}] \parallel
s_{1} \distributor [l_{13}]         \parallel
s_{2} \distributor [l_{23}]         \parallel
s_{3} \distributor [l_{30}]
\\
M_{\mathrm{o}} & =
l_{01} \distributor [r_{1}, l_{13}]         \parallel
l_{02} \distributor [r_{2}, l_{23}]         \parallel
l_{13} \distributor [r_{3}, l_{30}]         \parallel
l_{23} \distributor [r_{3}, l_{30}]         \parallel
l_{30} \distributor [r_{0}, l_{01}, l_{02}]
\end{align}

\section{Loss-Agnostic Behavioral Equivalence}

\label{section:behavioral-equivalence}

It is self-suggesting to consider weak bisimilarity as the kind of
equivalence that should hold between the two network processes
introduced in the previous section. Weak bisimilarity is a
well-established notion of behavioral equivalence that provides a
fine-grained distinction of observable behavior. Usually, weak
bisimilarity is a congruence with respect to most, if not all, process
constructors, allowing for modular reasoning. In the case of the
communication language, it is a congruence with respect to both parallel
composition and restriction.

The notion of behavior behind weak bisimilarity captures possible future
actions that are observable. The behavior of a process is essentially a
rooted graph where nodes are states, with the root node being the
current state, edges are state transitions, which are marked with the
observable actions that cause these transitions, and branching
characterizes non-determinism. In the case of the \TH-calculus and thus
the communication language, observable actions are send and receive
actions that involve global channels. With the expressivity of our
communication language, which is limited compared to that of the
\TH-calculus, behavior characterizes essentially how handing over
packets to global channels may result in packets appearing on possibly
other global channels.

Unfortunately, weak bisimilarity turns out to be too strict for our
situation, as it is able to distinguish between broadcast via multicast
and direct broadcast. To see why, assume in each of the networks shown
in Fig.~\ref{figure:broadcast} a packet is sent by node~$0$ and this
packet makes it to node~$3$. With direct broadcast, it is possible that
neither node~$1$ nor node~$2$ receives the packet as well. With
broadcast via multicast, however, node~$1$ or node~$2$ must receive it
and must do so before node~$3$ receives it.

This mismatch is an example of a more general issue, which shows up with
other sender--receiver pairs and also with other pairs of networks: With
direct broadcast, any arrival pattern is possible. With broadcast via
multicast, only arrival patterns that correspond in some sense to the
network topology are possible; more precisely, when a packet makes it to
some node, a path from the sender to the receiver must exist whose
intermediate nodes all receive the packet in the order they appear on
this path.

To remove this constraint on arrival patterns, we make the receive
channels lossy. This way, intermediate nodes are no longer guaranteed to
receive packets. It is not sufficient, however, to introduce this
lossiness for broadcast via multicast only; we need to introduce it also
for direct broadcast. This is because packet loss in the receive
channels is observable and consequently unilateral introduction of such
loss would create another behavioral mismatch between the two networks.

The approach of making receive channels lossy leads to a notion of
\emph{weak bisimilarity up to loss}, which is derived from weak
bisimilarity~(written $\approx$ in the following):
\begin{definition}[Weak bisimilarity up to loss]
Two processes $P$ and~$Q$ are weakly bisimilar up to loss in channels
$r_{1}$ to~$r_{n}$ exactly if
\begin{equation*}
\loser r_{1} \parallel \ldots \parallel \loser r_{n} \parallel P
\approx
\loser r_{1} \parallel \ldots \parallel \loser r_{n} \parallel Q
\enspace.
\end{equation*}
\end{definition}

Note that describing the equivalence of the two broadcast networks using
weak bisimilarity up to loss does not result in a trivial statement. In
particular, weak bisimilarity up to loss does not just consider the
situation where all packets are lost. Instead, it considers also
non-trivial arrival patterns, because the losers that are attached to
the receive channels compete with the environment for packets, so that
packets may escape the grip of the losers.

\section{A Proof of Correctness of Broadcast via Multicast}

\label{section:correctness-proof}

Broadcast via multicast is expected to behave equivalently to direct
broadcast as long as the multicast network is strongly connected. In
this work, however, we prove this equivalence only for the particular
networks depicted in Fig.~\ref{figure:broadcast}, which nevertheless
capture important phenomena that show up in other cases:
\begin{itemize}

\item

The multicast network has a node with several outgoing and a node with
several incoming links.

\item

In the multicast network, some nodes are reachable from certain other
nodes only via more than one hop.

\end{itemize}
Nothing in our proof is fundamentally tied to these particular networks,
though, and generalizing this proof to arbitrary pairs of a
broadcast-via-multicast network and a corresponding direct-broadcast
network should be straightforward. We merely chose a concrete example
for this early work on broadcast network equivalence to make the proof
easier to conduct.

Our concrete goal is to prove that $M$, defined in
(\ref{equation:broadcast-via-multicast}), and $D$, defined in
(\ref{equation:direct-broadcast}), are weakly bisimilar up to loss in
the receive channels, which is expressed by the statement
\begin{equation*}
\loser r_{0} \parallel
\loser r_{1} \parallel
\loser r_{2} \parallel
\loser r_{3} \parallel
M
\approx
\loser r_{0} \parallel
\loser r_{1} \parallel
\loser r_{2} \parallel
\loser r_{3} \parallel
D
\enspace.
\end{equation*}
We prove this statement by turning its left-hand side into its
right-hand side through a series of transformation steps, each of which
replaces subprocesses with bisimilar ones. These replacements are
justified by several basic lemmas about the communication language. Due
to time constraints, we have not yet proved all of these lemmas, but we
have reasonable confidence that they are correct.

\subsection{Correctness of Broadcast via Multicast: Graphically}

To outline our proof, we present the individual transformation steps and
the key lemmas used by them graphically using communication nets. The
lemmas are all bisimilarities, which are depicted in
Fig.~\ref{figure:key-bisimilarities}. The transformation steps are the
following ones:
\begin{enumerate}

\item

Untangling of receiving and relaying, which turns the
broadcast-via-multicast process depicted in Fig.~\ref{figure:broadcast}
with the receive channels made lossy into the process depicted in
Fig.~\ref{figure:after-untangling}, using the \textit{distributor-split}
lemma

\item

Transforming the core, which turns the result of the previous step into
the process depicted in Fig.~\ref{figure:after-core-transformation},
using the \textit{bridge-shortcut-redundancy} lemma

\item

Collapsing the sending part, which turns the result of the previous step
into the process depicted in Fig.~\ref{figure:after-sending-collapse},
using the \textit{distributor-target-fusion} lemma

\item

Collapsing the receiving part, which turns the result of the previous
step into the process depicted in
Fig.~\ref{figure:after-receiving-collapse}, using the
\textit{bridge-source-switch} lemma

\item

Collapsing the core, which turns the result of the previous step into
the direct-broadcast process depicted in Fig.~\ref{figure:broadcast}
with the receive channels made lossy, using the
\textit{duploss-detour-collapse} lemma

\end{enumerate}
\begin{figure}

\centering
\begin{tikzpicture}[communication net]

\begin{scope}

\node [place] (a)  at (-3,1) [label=left:$a$]      {\duplicating};
\node [place] (b1) at (-1,2) [label=right:$b_{1}$] {\losing};
\node [place] (bn) at (-1,0) [label=right:$b_{n}$] {\losing};

\node [symbol] at (-1,1) {\verticaldots};

\node [transition] at (-2,1) {} edge [pre]  (a)
                                edge [post] (b1)
                                edge [post] (bn);

\end{scope}

\node [symbol] at (0,1) {$\approx$};

\begin{scope}

\node [place] (a)  at (1,1) [label=left:$a$]      {\duplicating};
\node [place] (b1) at (3,2) [label=right:$b_{1}$] {\losing};
\node [place] (bn) at (3,0) [label=right:$b_{n}$] {\losing};

\node [transition] at (2,2) {} edge [pre]  (a)
                               edge [post] (b1);
\node [transition] at (2,0) {} edge [pre]  (a)
                               edge [post] (bn);

\node [symbol] at (2.5,1) {\verticaldots};

\end{scope}

\begin{scope}[captions]

\node at (0,0) {\textit{distributor-split}};

\end{scope}

\begin{scope}[yshift=-2.7cm]

\begin{scope}[bend angle=20]

\node [place] (a) at (-5,0) [label=left:$a$]  {};
\node [place] (b) at (-3,0) [label=above:$b$] {};
\node [place] (c) at (-1,0) [label=right:$c$] {};

\node [transition] at (-4,0) {} edge [pre]            (a)
                                edge [post]           (b);
\node [transition] at (-2,0) {} edge [pre]            (b)
                                edge [post]           (c);
\node [transition] at (-3,1) {} edge [pre,bend right] (a)
                                edge [post,bend left] (c);

\end{scope}

\node [symbol] at (0,0) {$\approx$};

\begin{scope}

\node [place] (a) at (1,0) [label=left:$a$]  {};
\node [place] (b) at (3,0) [label=above:$b$] {};
\node [place] (c) at (5,0) [label=right:$c$] {};

\node [transition] at (2,0) {} edge [pre]  (a)
                               edge [post] (b);
\node [transition] at (4,0) {} edge [pre]  (b)
                               edge [post] (c);

\end{scope}

\begin{scope}[captions]

\node at (0,0) {\textit{bridge-shortcut-redundancy}};

\end{scope}

\end{scope}

\begin{scope}[yshift=-6.4cm]

\begin{scope}[bend angle=20]

\node [place] (a)  at (-5,1) [label=left:$a$]     {};
\node [place] (b1) at (-3,2) [label=left:$b_{1}$] {};
\node [place] (bn) at (-3,0) [label=left:$b_{n}$] {};
\node [place] (c)  at (-1,1) [label=right:$c$]    {\duplosing};

\node [transition] at (-4,  1) {} edge [pre]            (a)
                                  edge [post]           (b1)
                                  edge [post]           (bn);
\node [transition] at (-2,  2) {} edge [pre]            (b1)
                                  edge [post,bend left] (c);
\node [transition] at (-2,1.5) {} edge [pre]            (c)
                                  edge [post]           (b1);
\node [transition] at (-2,0.5) {} edge [pre]            (bn)
                                  edge [post]           (c);
\node [transition] at (-2,  0) {} edge [pre,bend right] (c)
                                  edge [post]           (bn);

\node [symbol] at (-2.5,1) {\verticaldots};

\end{scope}

\node [symbol] at (0,1) {$\approx$};

\begin{scope}[bend angle=20]

\node [place] (a)  at (1,1) [label=left:$a$]      {};
\node [place] (b1) at (5,2) [label=right:$b_{1}$] {};
\node [place] (bn) at (5,0) [label=right:$b_{n}$] {};
\node [place] (c)  at (3,1) [label=right:$c$]     {\duplosing};

\node [transition] at (2,  1) {} edge [pre]            (a)
                                 edge [post]           (c);
\node [transition] at (4,  2) {} edge [pre,bend right] (c)
                                 edge [post]           (b1);
\node [transition] at (4,1.5) {} edge [pre]            (b1)
                                 edge [post]           (c);
\node [transition] at (4,0.5) {} edge [pre]            (c)
                                 edge [post]           (bn);
\node [transition] at (4,  0) {} edge [pre]            (bn)
                                 edge [post,bend left] (c);

\node [symbol] at (4.5,1) {\verticaldots};

\end{scope}

\begin{scope}[captions]

\node at (0,0) {\textit{distributor-target-fusion}};

\end{scope}

\end{scope}

\begin{scope}[yshift=-10.1cm]

\begin{scope}[bend angle=20]

\node [place] (a1) at (-3.25,2) [label=left:$a_{1}$] {};
\node [place] (a2) at (-3.25,0) [label=left:$a_{2}$] {};
\node [place] (b)  at (   -1,1) [label=right:$b$]    {};

\node [transition] at (  -2,1) {} edge [pre,bend left] (a1)
                                  edge [post]          (b);
\node [transition] at (  -3,1) {} edge [pre]           (a1)
                                  edge [post]          (a2);
\node [transition] at (-3.5,1) {} edge [pre]           (a2)
                                  edge [post]          (a1);

\end{scope}

\node [symbol] at (0,1) {$\approx$};

\begin{scope}[bend angle=20]

\node [place] (a1) at (   1,2) [label=left:$a_{1}$] {};
\node [place] (a2) at (   1,0) [label=left:$a_{2}$] {};
\node [place] (b)  at (3.25,1) [label=right:$b$]    {};

\node [transition] at (2.25,1) {} edge [pre,bend right] (a2)
                                  edge [post]           (b);
\node [transition] at (1.25,1) {} edge [pre]            (a1)
                                  edge [post]           (a2);
\node [transition] at (0.75,1) {} edge [pre]            (a2)
                                  edge [post]           (a1);

\end{scope}

\begin{scope}[captions]

\node at (0,0) {\textit{bridge-source-switch}};

\end{scope}

\end{scope}

\begin{scope}[yshift=-12.3cm]

\begin{scope}

\node [place] (a) at (-3,0.25) [label=left:$a$]        {};
\node [place] (b) at (-1,0.25) [local,label=right:$b$] {\duplosing};

\node [transition] at (-2,0.5) {} edge [pre]  (a)
                                  edge [post] (b);
\node [transition] at (-2,  0) {} edge [pre]  (b)
                                  edge [post] (a);

\end{scope}

\node [symbol] at (0,0.25) {$\approx$};

\begin{scope}

\node [place] (a) at (1,0.25) [label=left:$a$] {\duplosing};

\end{scope}

\begin{scope}[captions]

\node at (0,0) {\textit{duploss-detour-collapse}};

\end{scope}

\end{scope}

\end{tikzpicture}

\caption{The key bisimilarities used by the proof}

\label{figure:key-bisimilarities}

\end{figure}
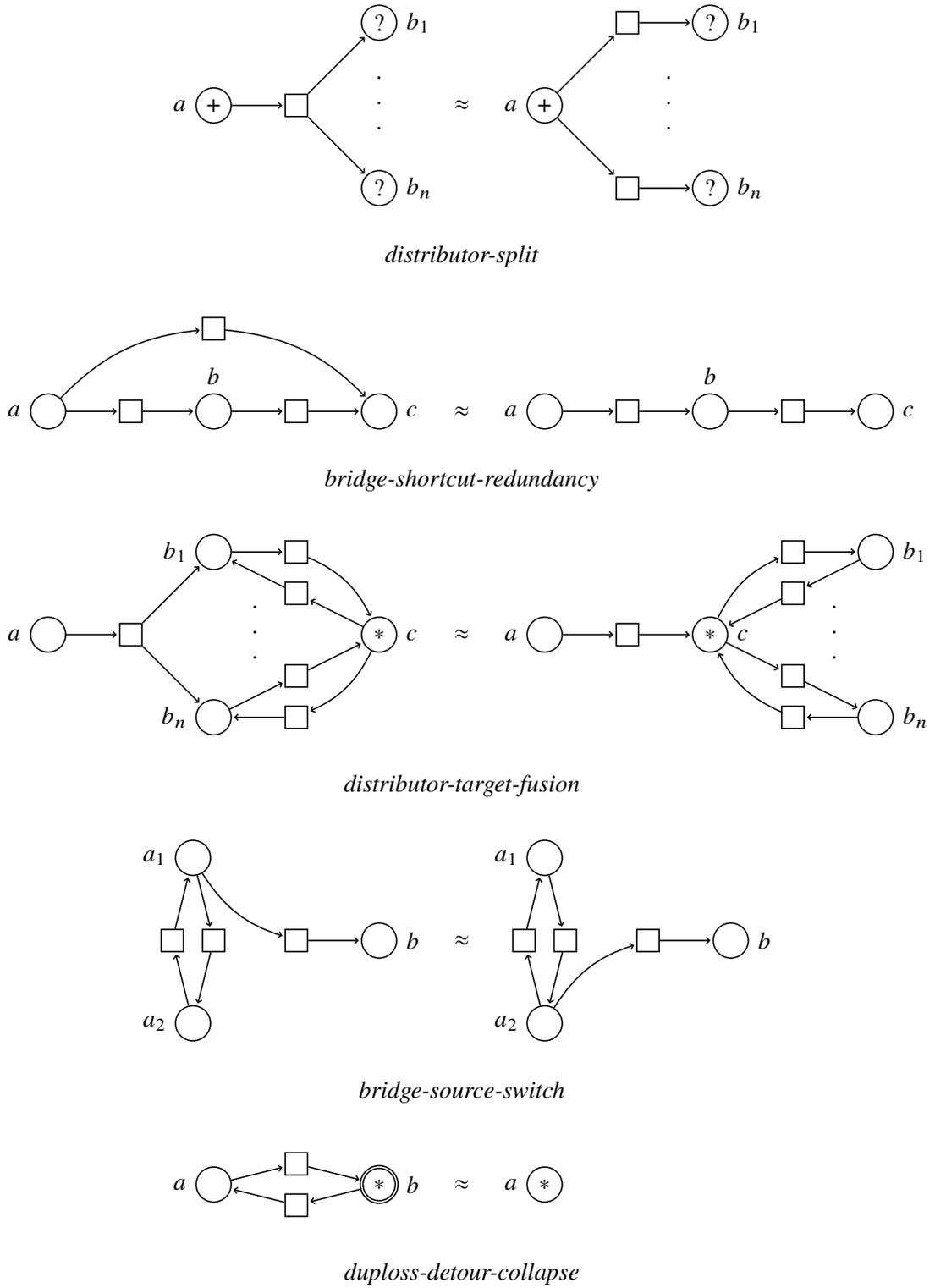
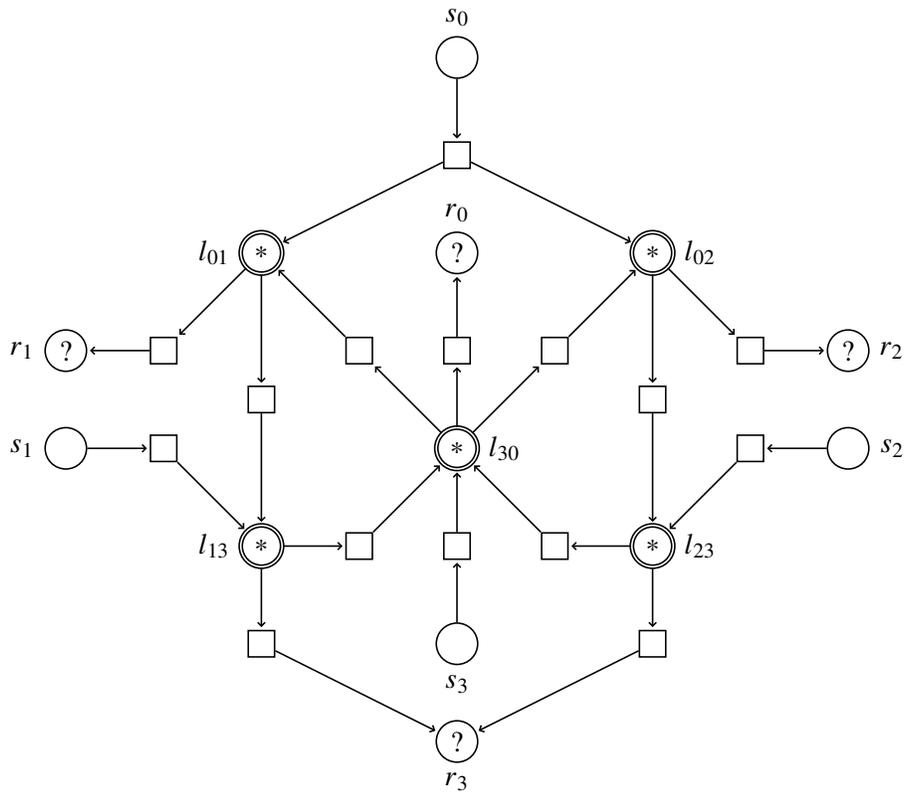
\begin{figure}

\centering
\begin{tikzpicture}[communication net]

\newcommand{\ls}{\losing}
\newcommand{\dl}{\duplosing}

\node [place] (s0)  at ( 0, 4) [label=above:$s_{0}$]        {};
\node [place] (s1)  at (-4, 0) [label=left:$s_{1}$]         {};
\node [place] (s2)  at ( 4, 0) [label=right:$s_{2}$]        {};
\node [place] (s3)  at ( 0,-2) [label=below:$s_{3}$]        {};
\node [place] (r0)  at ( 0, 2) [label=above:$r_{0}$]        {\ls};
\node [place] (r1)  at (-4, 1) [label=left:$r_{1}$]         {\ls};
\node [place] (r2)  at ( 4, 1) [label=right:$r_{2}$]        {\ls};
\node [place] (r3)  at ( 0,-3) [label=below:$r_{3}$]        {\ls};
\node [place] (l01) at (-2, 2) [local,label=left:$l_{01}$]  {\dl};
\node [place] (l02) at ( 2, 2) [local,label=right:$l_{02}$] {\dl};
\node [place] (l13) at (-2,-1) [local,label=left:$l_{13}$]  {\dl};
\node [place] (l23) at ( 2,-1) [local,label=right:$l_{23}$] {\dl};
\node [place] (l30) at ( 0, 0) [local,label=right:$l_{30}$] {\dl};

\node [transition] at ( 0,  3) {} edge [pre]  (s0)
                                  edge [post] (l01)
                                  edge [post] (l02);
\node [transition] at (-3,  0) {} edge [pre]  (s1)
                                  edge [post] (l13);
\node [transition] at ( 3,  0) {} edge [pre]  (s2)
                                  edge [post] (l23);
\node [transition] at ( 0, -1) {} edge [pre]  (s3)
                                  edge [post] (l30);
\node [transition] at (-3,  1) {} edge [pre]  (l01)
                                  edge [post] (r1);
\node [transition] at ( 3,  1) {} edge [pre]  (l02)
                                  edge [post] (r2);
\node [transition] at (-2, -2) {} edge [pre]  (l13)
                                  edge [post] (r3);
\node [transition] at ( 2, -2) {} edge [pre]  (l23)
                                  edge [post] (r3);
\node [transition] at ( 0,  1) {} edge [pre]  (l30)
                                  edge [post] (r0);
\node [transition] at (-2,0.5) {} edge [pre]  (l01)
                                  edge [post] (l13);
\node [transition] at ( 2,0.5) {} edge [pre]  (l02)
                                  edge [post] (l23);
\node [transition] at (-1, -1) {} edge [pre]  (l13)
                                  edge [post] (l30);
\node [transition] at ( 1, -1) {} edge [pre]  (l23)
                                  edge [post] (l30);
\node [transition] at (-1,  1) {} edge [pre]  (l30)
                                  edge [post] (l01);
\node [transition] at ( 1,  1) {} edge [pre]  (l30)
                                  edge [post] (l02);

\end{tikzpicture}

\caption{The process after untangling of receiving and relaying}

\label{figure:after-untangling}

\end{figure}
\begin{figure}

\centering
\begin{tikzpicture}[communication net]

\newcommand{\ls}{\losing}
\newcommand{\dl}{\duplosing}
\newcommand{\m}{$l_{30}$}

\node [place] (s0)  at ( 0, 4) [label=above:$s_{0}$]              {};
\node [place] (s1)  at (-4, 0) [label=left:$s_{1}$]               {};
\node [place] (s2)  at ( 4, 0) [label=right:$s_{2}$]              {};
\node [place] (s3)  at ( 0,-2) [label=below:$s_{3}$]              {};
\node [place] (r0)  at ( 0, 2) [label=above:$r_{0}$]              {\ls};
\node [place] (r1)  at (-4, 1) [label=left:$r_{1}$]               {\ls};
\node [place] (r2)  at ( 4, 1) [label=right:$r_{2}$]              {\ls};
\node [place] (r3)  at ( 0,-3) [label=below:$r_{3}$]              {\ls};
\node [place] (l01) at (-2, 2) [local,label=left:$l_{01}$]        {\dl};
\node [place] (l02) at ( 2, 2) [local,label=right:$l_{02}$]       {\dl};
\node [place] (l13) at (-2,-1) [local,label=left:$l_{13}$]        {\dl};
\node [place] (l23) at ( 2,-1) [local,label=right:$l_{23}$]       {\dl};
\node [place] (l30) at ( 0, 0) [local,label={[anchor=west]15:\m}] {\dl};

\node [transition] at ( 0,    3) {} edge [pre]  (s0)
                                    edge [post] (l01)
                                    edge [post] (l02);
\node [transition] at (-3,    0) {} edge [pre]  (s1)
                                    edge [post] (l13);
\node [transition] at ( 3,    0) {} edge [pre]  (s2)
                                    edge [post] (l23);
\node [transition] at ( 0,   -1) {} edge [pre]  (s3)
                                    edge [post] (l30);
\node [transition] at (-3,    1) {} edge [pre]  (l01)
                                    edge [post] (r1);
\node [transition] at ( 3,    1) {} edge [pre]  (l02)
                                    edge [post] (r2);
\node [transition] at (-2,   -2) {} edge [pre]  (l13)
                                    edge [post] (r3);
\node [transition] at ( 2,   -2) {} edge [pre]  (l23)
                                    edge [post] (r3);
\node [transition] at ( 0,    1) {} edge [pre]  (l30)
                                    edge [post] (r0);
\node [transition] at (-1,-0.25) {} edge [pre]  (l13)
                                    edge [post] (l30);
\node [transition] at (-1,-0.75) {} edge [pre]  (l30)
                                    edge [post] (l13);
\node [transition] at ( 1,-0.25) {} edge [pre]  (l30)
                                    edge [post] (l23);
\node [transition] at ( 1,-0.75) {} edge [pre]  (l23)
                                    edge [post] (l30);
\node [transition] at (-1, 1.25) {} edge [pre]  (l01)
                                    edge [post] (l30);
\node [transition] at (-1, 0.75) {} edge [pre]  (l30)
                                    edge [post] (l01);
\node [transition] at ( 1, 1.25) {} edge [pre]  (l30)
                                    edge [post] (l02);
\node [transition] at ( 1, 0.75) {} edge [pre]  (l02)
                                    edge [post] (l30);

\end{tikzpicture}

\caption{The process after transforming the core}

\label{figure:after-core-transformation}

\end{figure}
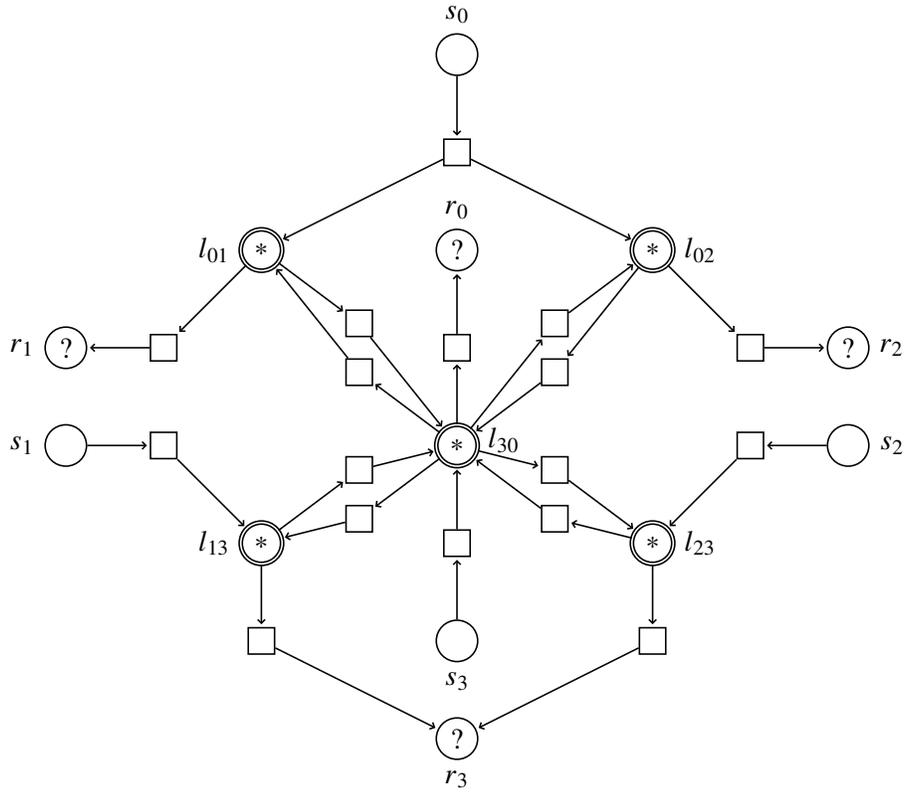
\begin{figure}

\centering
\begin{tikzpicture}[communication net]

\newcommand{\ls}{\losing}
\newcommand{\dl}{\duplosing}

\node [place] (s0)  at ( 0.5,   2) [label=above:$s_{0}$]        {};
\node [place] (s1)  at (  -4,-0.5) [label=left:$s_{1}$]         {};
\node [place] (s2)  at (   4,-0.5) [label=right:$s_{2}$]        {};
\node [place] (s3)  at (   0,  -2) [label=below:$s_{3}$]        {};
\node [place] (r0)  at (-0.5,   2) [label=above:$r_{0}$]        {\ls};
\node [place] (r1)  at (  -4, 0.5) [label=left:$r_{1}$]         {\ls};
\node [place] (r2)  at (   4, 0.5) [label=right:$r_{2}$]        {\ls};
\node [place] (r3)  at (   0,  -3) [label=below:$r_{3}$]        {\ls};
\node [place] (l01) at (  -2, 0.5) [local,label=above:$l_{01}$] {\dl};
\node [place] (l02) at (   2, 0.5) [local,label=above:$l_{02}$] {\dl};
\node [place] (l13) at (  -1,  -2) [local,label=left:$l_{13}$]  {\dl};
\node [place] (l23) at (   1,  -2) [local,label=right:$l_{23}$] {\dl};
\node [place] (l30) at (   0,   0) [local,label=above:$l_{30}$] {\dl};

\node [transition] at ( 0.5,   1) {} edge [pre]  (s0)
                                     edge [post] (l30);
\node [transition] at (  -1,-0.5) {} edge [pre]  (s1)
                                     edge [post] (l30);
\node [transition] at (   1,-0.5) {} edge [pre]  (s2)
                                     edge [post] (l30);
\node [transition] at (   0,  -1) {} edge [pre]  (s3)
                                     edge [post] (l30);
\node [transition] at (  -3, 0.5) {} edge [pre]  (l01)
                                     edge [post] (r1);
\node [transition] at (   3, 0.5) {} edge [pre]  (l02)
                                     edge [post] (r2);
\node [transition] at (  -1,  -3) {} edge [pre]  (l13)
                                     edge [post] (r3);
\node [transition] at (   1,  -3) {} edge [pre]  (l23)
                                     edge [post] (r3);
\node [transition] at (-0.5,   1) {} edge [pre]  (l30)
                                     edge [post] (r0);
\node [transition] at (  -1,  -1) {} edge [pre]  (l13)
                                     edge [post] (l30);
\node [transition] at (-0.5,  -1) {} edge [pre]  (l30)
                                     edge [post] (l13);
\node [transition] at (   1,  -1) {} edge [pre]  (l30)
                                     edge [post] (l23);
\node [transition] at ( 0.5,  -1) {} edge [pre]  (l23)
                                     edge [post] (l30);
\node [transition] at (  -1, 0.5) {} edge [pre]  (l01)
                                     edge [post] (l30);
\node [transition] at (  -1,   0) {} edge [pre]  (l30)
                                     edge [post] (l01);
\node [transition] at (   1, 0.5) {} edge [pre]  (l30)
                                     edge [post] (l02);
\node [transition] at (   1,   0) {} edge [pre]  (l02)
                                     edge [post] (l30);

\end{tikzpicture}

\caption{The process after collapsing the sending part}

\label{figure:after-sending-collapse}

\end{figure}
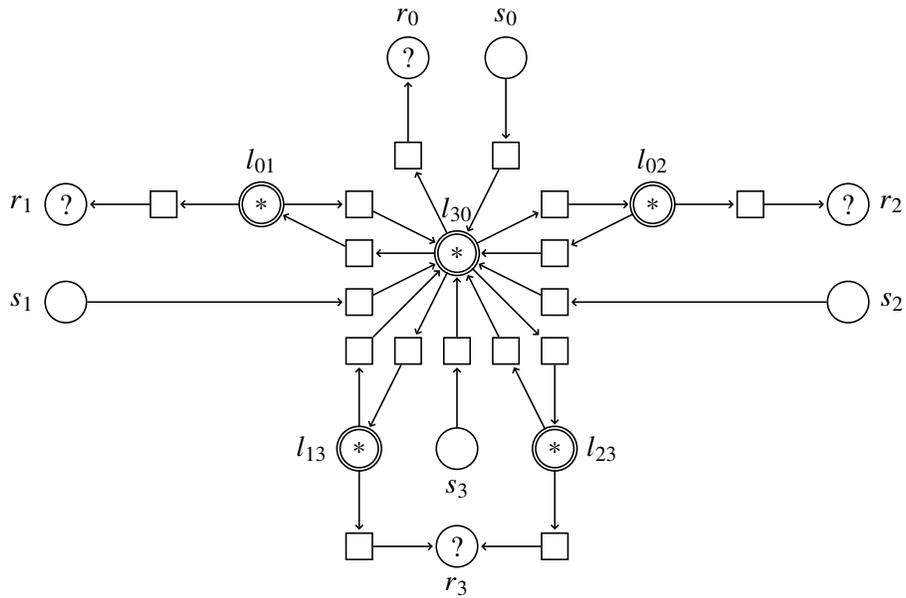
\begin{figure}

\centering
\begin{tikzpicture}[communication net]

\newcommand{\ls}{\losing}
\newcommand{\dl}{\duplosing}

\node [place] (s0)  at ( 0.5,   2) [label=above:$s_{0}$]        {};
\node [place] (s1)  at (  -3,-0.5) [label=left:$s_{1}$]         {};
\node [place] (s2)  at (   3,-0.5) [label=right:$s_{2}$]        {};
\node [place] (s3)  at ( 0.5,  -2) [label=below:$s_{3}$]        {};
\node [place] (r0)  at (-0.5,   2) [label=above:$r_{0}$]        {\ls};
\node [place] (r1)  at (  -3, 0.5) [label=left:$r_{1}$]         {\ls};
\node [place] (r2)  at (   3, 0.5) [label=right:$r_{2}$]        {\ls};
\node [place] (r3)  at (-0.5,  -2) [label=below:$r_{3}$]        {\ls};
\node [place] (l01) at (  -2,   1) [local,label=above:$l_{01}$] {\dl};
\node [place] (l02) at (   2,   1) [local,label=above:$l_{02}$] {\dl};
\node [place] (l13) at (  -2,  -1) [local,label=left:$l_{13}$]  {\dl};
\node [place] (l23) at (   2,  -1) [local,label=right:$l_{23}$] {\dl};
\node [place] (l30) at (   0,   0) [local,label=above:$l_{30}$] {\dl};

\node [transition] at ( 0.5,    1) {} edge [pre]  (s0)
                                      edge [post] (l30);
\node [transition] at (  -2,-0.25) {} edge [pre]  (s1)
                                      edge [post] (l30);
\node [transition] at (   2,-0.25) {} edge [pre]  (s2)
                                      edge [post] (l30);
\node [transition] at ( 0.5,   -1) {} edge [pre]  (s3)
                                      edge [post] (l30);
\node [transition] at (  -2, 0.25) {} edge [pre]  (l30)
                                      edge [post] (r1);
\node [transition] at (   2, 0.25) {} edge [pre]  (l30)
                                      edge [post] (r2);
\node [transition] at (-0.5,   -1) {} edge [pre]  (l30)
                                      edge [post] (r3);
\node [transition] at (-0.5,    1) {} edge [pre]  (l30)
                                      edge [post] (r0);
\node [transition] at (  -1, -0.5) {} edge [pre]  (l13)
                                      edge [post] (l30);
\node [transition] at (  -1,   -1) {} edge [pre]  (l30)
                                      edge [post] (l13);
\node [transition] at (   1, -0.5) {} edge [pre]  (l30)
                                      edge [post] (l23);
\node [transition] at (   1,   -1) {} edge [pre]  (l23)
                                      edge [post] (l30);
\node [transition] at (  -1,    1) {} edge [pre]  (l01)
                                      edge [post] (l30);
\node [transition] at (  -1,  0.5) {} edge [pre]  (l30)
                                      edge [post] (l01);
\node [transition] at (   1,    1) {} edge [pre]  (l30)
                                      edge [post] (l02);
\node [transition] at (   1,  0.5) {} edge [pre]  (l02)
                                      edge [post] (l30);

\end{tikzpicture}

\caption{The process after collapsing the receiving part}

\label{figure:after-receiving-collapse}

\end{figure}
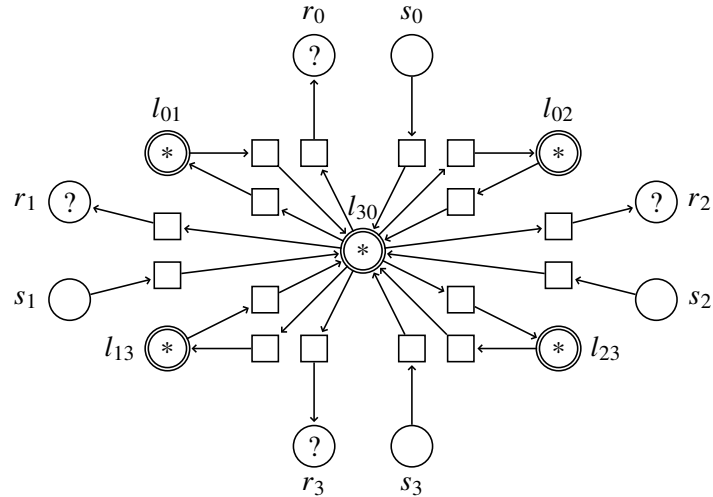

\subsection{Correctness of Broadcast via Multicast: Formally}

We have developed our proof formally using Isabelle/HOL.\footnote{See
\url{https://github.com/input-output-hk/network-equivalences}.} The
proof is phrased in a style similar to equational reasoning, with the
difference that we use weak bisimilarity instead of equality. Since
Isabelle does not come with support for automated rewriting based on
equivalence relations other than equality, we have developed a
corresponding extension,\footnote{See
\url{https://github.com/input-output-hk/equivalence-reasoner}.} which we
use in our proof.

While the formal proof closely follows its communication net
counterpart, it contains more technical detail. In particular, it
includes applications of those bisimilarities that are implicit in the
communication net notation, as described in
Sect.~\ref{section:communication-language}. However, our engine for
equivalence-based rewriting is able to automatically find any proof that
consists of a chain of rewriting steps involving only these
bisimilarities and optionally certain idempotency statements, which are
bisimilarities as well. As a result, we can bundle consecutive
applications of all these bisimilarities in our formal proof, which
therefore is still reasonably concise and readable.

To illustrate the general style of the proof, we show how the first
transformation step is conducted. We do not use Isabelle syntax for this
presentation, in order to make it slightly more accessible. However, the
Isabelle version is almost identical to what we show here, the main
difference being that it mentions the repeated use of the equivalence
reasoner and the lemmas to use as rewrite rules. In the proof snippet,
we use the notation $\Parallel{a}{[b_{1}, \ldots, b_{n}]} P$, which
stands for $P[b_{1}/a] \parallel \ldots \parallel P[b_{n}/a]$ and has
higher precedence than~$\parallel$.

The first transformation step consists of three substeps, which are
presented in Fig.~\ref{figure:first-transformation-step}. The second
substep is where the \textit{distributor-split} lemma is invoked. The
first and the third substep apply bisimilarities implicit in the
communication net notation and idempotency statements: the first substep
does so to enable the invocation of the \textit{distributor-split}
lemma; the third substep does so to undo modifications done by the first
substep and further streamline the term.
\begin{figure}

\centering
\begin{math}
\newcommand{\and}{\parallel\\&\mathrel{\phantom{\approx}}}
\begin{aligned}
\loser r_{0} \parallel
\loser r_{1} \parallel
\loser r_{2} \parallel
\loser r_{3} \parallel
M_{*} \parallel M_{\mathrm{o}}
& \approx
(
    \duplicator l_{01}
    \parallel \Parallel{a}{[r_{1}, l_{13}]} \loser a
    \parallel l_{01} \distributor [r_{1}, l_{13}]
)
\and
(
    \duplicator l_{02}
    \parallel \Parallel{a}{[r_{2}, l_{23}]} \loser a
    \parallel l_{02} \distributor [r_{2}, l_{23}]
)
\and
(
    \duplicator l_{13}
    \parallel \Parallel{a}{[r_{3}, l_{30}]} \loser a
    \parallel l_{13} \distributor [r_{3}, l_{30}]
)
\and
(
    \duplicator l_{23}
    \parallel \Parallel{a}{[r_{3}, l_{30}]} \loser a
    \parallel l_{23} \distributor [r_{3}, l_{30}]
)
\and
(
    \duplicator l_{30}
    \parallel \Parallel{a}{[r_{0}, l_{01}, l_{02}]} \loser a
    \parallel l_{30} \distributor [r_{0}, l_{01}, l_{02}]
)
\\
& \approx
(
    \duplicator l_{01}
    \parallel \Parallel{a}{[r_{1}, l_{13}]} \loser a
    \parallel \Parallel{a}{[r_{1}, l_{13}]} l_{01} \bridge a
)
\and
(
    \duplicator l_{02}
    \parallel \Parallel{a}{[r_{2}, l_{23}]} \loser a
    \parallel \Parallel{a}{[r_{2}, l_{23}]} l_{02} \bridge a
)
\and
(
    \duplicator l_{13}
    \parallel \Parallel{a}{[r_{3}, l_{30}]} \loser a
    \parallel \Parallel{a}{[r_{3}, l_{30}]} l_{13} \bridge a
)
\and
(
    \duplicator l_{23}
    \parallel \Parallel{a}{[r_{3}, l_{30}]} \loser a
    \parallel \Parallel{a}{[r_{3}, l_{30}]} l_{23} \bridge a
)
\and
(
    \duplicator l_{30}
    \parallel \Parallel{a}{[r_{0}, l_{01}, l_{02}]} \loser a
    \parallel \Parallel{a}{[r_{0}, l_{01}, l_{02}]} l_{30} \bridge a
)
\\
& \approx
\loser r_{0} \parallel
\loser r_{1} \parallel
\loser r_{2} \parallel
\loser r_{3} \parallel
M_{*} \and
l_{01} \bridge r_{1} \parallel
l_{02} \bridge r_{2} \parallel
l_{13} \bridge r_{3} \parallel
l_{23} \bridge r_{3} \parallel
l_{30} \bridge r_{0} \and
l_{01} \bridge l_{13} \parallel
l_{02} \bridge l_{23} \parallel
l_{13} \bridge l_{30} \parallel
l_{23} \bridge l_{30} \parallel
l_{30} \bridge l_{01} \parallel
l_{30} \bridge l_{02}
\end{aligned}
\end{math}

\caption{The first transformation step in detail}

\label{figure:first-transformation-step}

\end{figure}

\section{Related Work}

\label{section:related-work}

The correctness of broadcast via multicast, commonly referred to as
\emph{network flooding}, and similar techniques has been studied to some
extent in the literature. In the following, we discuss two
characteristic examples:
\begin{itemize}

\item

Bani-Abdelrahman~\cite{bani-abdelrahman:2018} formally specifies
synchronous and bounded asynchronous flooding algorithms using LTL and
verifies them using the model checker NuSMV. His results are limited to
small network sizes and fixed delays, though.

\item

Bar-Yehuda et~al.~\cite{bar-yehuda:1989} emulate a single-hop
(direct-broadcast) network with a multi-hop network using a synchronous
gossiping algorithm. With a gossiping algorithm, a node does not relay
an incoming packet to all neighboring nodes, but only to a randomly
chosen one.

\end{itemize}
The existing literature, however, seems to lack the study of an
\emph{equivalence} of the two broadcasting approaches, which we provide
in this work. The reason for this lack may be related to the
complications that arise due to the behavioral mismatches explained in
Sect.~\ref{section:behavioral-equivalence}.

There is a vast amount of literature on the relationship between process
calculi and Petri nets. In particular, a long line of research has been
developed with the theme of giving Petri net semantics to process
calculi~\cite{best:2001,degano:1988,olderog:1991}, providing process
calculi with operational semantics expressing true concurrency as
opposed to the traditional interleaving semantics. Another line of
research approaches the reverse problem, that is, finding process
calculi that are suitable for modeling Petri nets of certain
classes~\cite{basten:1995,gorrieri:2017}. Our tandem of the
communication language and its communication net notation, described in
Sect.~\ref{section:communication-language}, fits right into all this
research: the communication language can be regarded as a restricted
process calculus that comes with a Petri net semantics and models the
class of hierarchical Petri nets with exactly one input place per
transition.

\section{Conclusion}

\label{section:conclusion}

We have defined a language for describing communication networks, which
is embedded in a process calculus and corresponds to a class of Petri
nets. Based on the connection to Petri nets, we have devised a graphical
notation for our language. Building on this foundation, we have proved
behavioral equivalence of broadcast via multicast and direct broadcast
for a typical pair of networks. The graphical notation has allowed us to
reason in an intuitive way on an abstract level, while the embedding in
a process calculus has permitted us to develop a fully machine-checked
proof. For specifying the equivalence of the two realizations of
broadcast, we have devised the notion of weak bisimilarity up to loss.

\section{Ongoing and Future Work}

\label{section:ongoing-and-future-work}

At the moment, we are completing the proofs of the fundamental lemmas
that the correctness proof shown in
Sect.~\ref{section:correctness-proof} uses. Furthermore, we are working
on a variant of our correctness proof that deals with broadcast
integrated with packet filtering according to a fixed predicate.

An important task for the future is the generalization of our
correctness proofs to apply to arbitrary strongly connected multicast
networks and their direct-broadcast counterparts. Furthermore, we may
prove a modified correctness statement where the receive channels of
direct broadcast are not lossy. Doing so would clarify that broadcast
via multicast has the more constrained behavior, but would require
adjustments to the specification of network behavior. Finally, we
consider generalizing the proof about broadcast integrated with
filtering to work with state-dependent filters.

\section*{Acknowledgements}

This work was funded by Input Output. We are thankful to Input Output
for giving us the opportunity to work on numerous interesting topics,
including the one described in this paper. Furthermore, we want to thank
James Chapman, Duncan Coutts, Kevin Hammond, and Philipp Kant, who
supported us in our work on network equivalences and the development of
all the theory and the vast amount of Isabelle formalizations that
underlie it.

\bibliographystyle{eptcs}
\bibliography{paper}

\end{document}